\documentclass[a4paper]{report}
\usepackage[utf8]{inputenc}
\usepackage[T1]{fontenc}
\usepackage{RJournal}
\usepackage{amsmath,amssymb,array}
\usepackage{booktabs}

\usepackage{bm}

\begin{document}

\sectionhead{Contributed research article}
\volume{XX}
\volnumber{YY}
\year{20ZZ}
\month{AAAA}

\begin{article}

\title{\pkg{ctsmr} -- Continuous Time Stochastic Modeling in R}
\author{by Rune Juhl, Jan Kloppenborg Møller and Henrik Madsen}

\maketitle

\abstract{
\pkg{ctsmr} is an R package providing a general framework for identifying and estimating partially observed continuous-discrete time gray-box models. The estimation is based on maximum likelihood principles and Kalman filtering efficiently implemented in Fortran. This paper briefly demonstrates how to construct a Continuous Time Stochastic Model using multivariate time series data, and how to estimate the embedded parameters. The setup provides a unique framework for statistical modeling of physical phenomena, and the approach is often called grey box modeling. Finally three examples are provided to demonstrate the capabilities of \pkg{ctsmr}.
}

\section{Introduction}

The \pkg{ctsmr} package is an R package for identifying and estimating
continuous-discrete time state space models. The state space model
consists of continuous time \emph{system equations} formulated using
Stochastic Differential Equations (SDEs), and discrete
time \emph{measurement equations} which describe how the measurements
relate to the states of the system. The state space formulation
quantifies the amount of system and measurement noise. The system
noise originates from model approximations and uncertainties in the
input variables, whereas the measurement noise is related solely to
uncertainties of the measurements.

This modeling approach bridges the gap between physical and
statistical modeling, and it facilitates a setup which exploits the 
prior physical knowledge about the system, but the model structure and
parameters are not assumed to be completely known. The suggested
modeling framework is often called grey box modeling, cf. \cite{bolin1994} and \cite{torn2004b}.

In fact this modeling framework opens up for new tools for model
development. Specifically the approach allows for tracking of unknown
inputs and parameters over time by modeling them as random walk
processes. These principles lead to efficient methods for pinpointing
model deficiencies, and subsequently for identifying model
improvements. The approach also provides methods for model validation.

Modeling in continuous time has its benefits. The observations may be sampled irregularly for additional flexibility during an experiment. The models can be linear or nonlinear, and stationary or nonstationary. The parameters are not a function of the discretization but rather a function of the natural physical dimension, i.e. seconds is time is in seconds. Thus interpretation of the model parameters is easier also for experts of the modeled physical system.

In this study we will use maximum likelihood techniques both for parameter estimation and for model identification and verification. Like in \cite{kristensen2004a} the Extended Kalman Filter is used for evaluating the likelihood function.

There are a number of packages for discrete time dynamic linear modeling: \CRANpkg{dlm} \citep{Petris:2010:JSSOBK:v36i12}. A few packages deal with continuous time models: \CRANpkg{cts} \citep{cts} is restricted to autoregressive models, \CRANpkg{sde} \citep{iacus:2014} and \CRANpkg{yuima} \citep{yuima} deals with a broad range of diffusion processes, i.e. fractal Brownian motions and jumps, but does not allow for partial observation of the states. We consider \pkg{ctsmr} to be a more general framework for gray-box models which are physical models fitted to data.

\pkg{ctsmr} has been successfully applied to a range of applications, e.g.: heat dynamics of thermal systems (walls and buildings \citep{Bacher2011}, building integrated photovoltaic systems \citep{lodi_modelling_2012}), solar and wind power forecasting \citep{iversen_probabilistic_2013}, solar-activity \citep{vio_stochastic_2006}, pharmacokinetic/pharmacodynamic \citep{hansen_predicting_2014} and rainfall-runoff forecasting \citep{lowe_stochastic_2014}.

This paper shortly introduces the concepts of \pkg{ctsmr} and the model class it deals with. \pkg{ctsmr} provides functions for computing various conditional estimates, log-likelihood, AIC, BIC and other useful measures. For updated information about the package, examples and the user's guide see \url{http://ctsm.info} and \citep{juhl:2015a}.

\section{The CTSM-R model class}
The model class consist of a set of system and measurement equations in a state space formulation. The physical system is described by the system equations (\ref{eq:state_sde}) which is a set of (continuous time) It\^o stochastic differential equations. The stochastic process $\bm x_t$ represents the state of the system. The state is not observed directly, but the discrete time measurement equation (\ref{eq:obs}) describes how the (possibly multivariate) time series of measurements $\bm y_k$ relates to the state. The continuous-discrete time state space model used in \pkg{ctsmr} is
\begin{align}
  d\bm{x}_t &= \bm{f}\left( \bm{x}_t, \bm{u}_t, t, \bm{\theta} \right) dt +  \bm{\sigma}\left(\bm{u}_t, t, \bm{\theta} \right) d\bm{\omega}_t \label{eq:state_sde} \\
  \bm{y}_k &= \bm{h}\left( \bm{x}_k, \bm{u}_k, t_k, \bm{\theta} \right) + \bm{e}_k \qquad \bm e_k \sim \mathcal N (0,\bm{S}(\bm{u}_k,t_k)) \ , \label{eq:obs}
\end{align}
where $\bm x_t$ is the state, $\bm u_t$ is an exogenous input and $\bm \theta$ the parameters of the model. $\bm f()$ and $\bm \sigma()$ are possibly non-linear functions called the drift and diffusion terms. $\bm \omega$ is the Brownian motion driving the stochastic part of the system equations. The measurement equation (\ref{eq:obs}) contains the function $\bm h()$ which is a possibly non-linear function of the states and inputs. The measurement noise $\bm e_k$ is independent from the diffusion in the system equations and models the imperfection of the measurements. $\bm{e}_k$ is Gaussian with $\mathcal N (0,\bm{S}(\bm{u}_k,t_k))$.

The general continuous-discrete time state space model (\ref{eq:state_sde})-(\ref{eq:obs}) is considered linear when both the system and measurement equations are linear in both states and inputs. The linear continuous-discrete time state space model used in \pkg{ctsmr} is
\begin{align}
  d\bm{x}_t &= \left( \bm{A}\left( t, \bm{\theta} \right)\bm{x}_t + \bm{B}\left( t, \bm{\theta} \right)\bm{u}_t \right) dt +  \bm{\sigma}\left(\bm{u}_t, t, \bm{\theta} \right) d\bm{\omega}_t \label{eq:linear_state_sde} \\
  \bm{y}_k &= \bm{C}\left( t_k, \bm{\theta} \right)\bm{x}_k + \bm{D}\left( t_k, \bm{\theta} \right)\bm{u}_k + \bm{e}_k \qquad \bm e_k \sim \mathcal N (0,\bm{S}(\bm{u}_k,t_k)) \ . \label{eq:linear_obs}
\end{align}

\subsection{Maximum likelihood}
The parameters are estimated by maximizing the likelihood function. Given a time series 
\begin{align}
   \mathcal{Y}_N = [\bm{y_0}, \bm{y_1}, \dots, \bm{y_k}, \dots, \bm{y}_N] \, 
\end{align}
the likelihood of the unknown parameters $\theta$ given the model formulated as (\ref{eq:state_sde})-(\ref{eq:obs}) is the joint probability density function (pdf)
\begin{align}
   L(\bm{\theta}) = \mathrm{p} \left( \mathcal{Y}_N \vert \bm{\theta} \right) = \left( \prod_{k=1}^{N} \mathrm{p} \left( \bm{y}_k \vert  \mathcal{Y}_{k-1}, \bm{\theta} \right) \right) \mathrm{p} (\bm{y}_0 \vert \bm{\theta}) \, , \label{eq:single_likelihood}
\end{align}
where the likelihood $L$ is the joint probability density function.

The solution to a linear SDE driven by a Brownian motion is a Gaussian process and the likelihood function is exact. Non-linear SDEs do not result in a Gaussian process and thus the marginal probability is not Gaussian. However, by sampling fast enough relative to the time constants of the system and how strong the non-linearities are, it is then reasonable to assume that the conditional density is approximately Gaussian. The conditional probability density of the states is thus described by the first and second order moments. The forward propagation of the mean and variance-covariance matrix is governed by two ODEs
\begin{align}
   \frac{d \hat{\bm{x}}_{t \vert k}}{dt} &= \bm{A}\hat{\bm{x}}_{t \vert k} + \bm{B}\bm{u}_t \, , \qquad t \in \left[ t_k, t_{k+1} \right] \\
   \frac{d \hat{\bm{P}}_{t \vert k}}{dt} &= \bm{A}\hat{\bm{P}}_{t \vert k} + \hat{\bm{P}}_{t \vert k}\bm{A}^{\prime} + \bm{\sigma}\bm{\sigma}^{\prime}\, , \qquad t \in \left[ t_k, t_{k+1} \right]
\end{align}
which are solved differently for linear and non-linear systems. The probability density function is then transformed through the measurement equation where the first and second order moments for the output is defined as
\begin{eqnarray}
   \hat{\bm{y}}_{k \vert k-1} &=& E \left[ \bm{y}_k \vert \mathcal{Y}_{k-1}, \bm{\theta} \right] \\
   \Sigma_{k \vert k-1} &=& V \left[ \bm{y}_k \vert \mathcal{Y}_{k-1}, \bm{\theta}  \right] \, .
\end{eqnarray}
Introducing the innovation error
\begin{align}
   \bm{ \varepsilon }_k = \bm{y}_k - \hat{\bm{y}}_{k \vert k-1} \, ,\label{eq:inovaerror}
\end{align}
the (approximate) likelihood (\ref{eq:single_likelihood}) becomes
\begin{align}
   L(\bm{\theta}, \mathcal{Y}_N) = \left( \prod_{k=1}^{N} \frac{ \mathrm{exp} \left( -\frac{1}{2} \varepsilon_k^T \Sigma_{k \vert k-1}^{-1} \varepsilon_k \right) }{\sqrt{ \vert \Sigma_{k \vert k-1} \vert } \sqrt{2 \pi}^l} \right) \mathrm{p} (\bm{y}_0 \vert \bm{\theta}) \, .
\end{align}
The probability density of the initial observation $p(y_0 \vert \bm{\theta})$ is parameterized through the probability density of the initial state $p(x_0 \vert \bm{\theta})$. The mean $\hat{\bm{y}}_{k \vert k-1}$ and covariance $\Sigma_{k \vert k-1}$ are computed recursively using either the standard Kalman filter for linear models or the extended Kalman filter for non-linear models, see \citep{jazwinski} for a detailed description.

The estimation of the unknown parameters is a non-linear optimization problem. The negative log-likelihood is minimized by a Quasi-Newton optimizer.
\begin{align}
   \hat{\bm{\theta}} = \arg \min_{\bm{\theta} \in \Theta} \left( -\mathrm{ln} \left( L(\bm{\theta}, \mathcal{Y}_N \right) \right) \, .
\end{align}

\pkg{ctsmr} can also be used in a Bayesian setting by specifying a Gaussian prior distribution of (possibly a subset) the parameters. This is results in the maximum a posterior (MAP) estimate. For details on how to use MAP in \pkg{ctsmr}, see \citep{juhl:2015a}. 

\section{Implementation}
The entire mathematical engine of \pkg{ctsmr} is written in \strong{FORTRAN} using \strong{BLAS} and \strong{LAPACK} basic mathematical operations. \pkg{ctsmr} comes with its own implementation of the standard and extended continuous-discrete Kalman filters using either \strong{Expokit} or numerical integrators \strong{ODEPACK} for the forward propagation of the ODEs for the mean and variance-covariance of the states. The calculation of the log-likelihood and its numerical gradient wrt. the parameters may be calculated in parallel on shared memory systems using \strong{OpenMP}. 

\pkg{ctsmr} comes with an interface written in \strong{R}. The matematical model is given using standard R formulaes. \pkg{ctsmr} manipulates the equations and determine if the model is linear or not. The model is then translated into \strong{FORTRAN} and compiled ensuring computational efficient calculations compared to within R.

\subsection{Model specification}
The model object is implemented using R's object oriented programming class \emph{reference classes}. The reason for using a reference class is that many models often have several system equations and/or measurement equations which makes the layered approach easy to get an overview of at the cost of compactness. Models can easily be extended by adding additional system or measurement equations. The interface can be accessed through loops thus enabling implementing a discretization stencil for a spatial model without having to specify every equation manually.

To build a model, an instance of the CTSM object has to be created.
\begin{example}
m <- ctsm()
\end{example}
Here \emph{m} is an empty model which must be manipulated through the methods attached to the object. The mathematical model can now be constructed using the following methods
\begin{itemize}
   \item \code{\$addSystem(sde\_formula)}
   \item \code{\$addObs(formula)}
   \item \code{\$setVariance(formula)}
   \item \code{\$addInput(name)}
\end{itemize}

\pkg{ctsmr} allows any mathematically correct SDE according to (\ref{eq:state_sde}) to be written as a standard R formula. To add a system equation to the model object, \emph{m}, use the \code{\$addSystem()} method
\begin{example}
m$addSystem(dx1 ~ (x1-x2)/C1 * dt + sigma * dw1)
\end{example}
From the given SDE \pkg{ctsmr} determines that \code{x1} is a state. Additional system equations may be added to the CTSM by additional use of \code{\$addSystem}.

The discrete time measurement equations are added by calling the \code{\$addObs} method. The specification of the equation is limited to the function $h(\cdot)$ of (\ref{eq:obs}).
\begin{example}
m$addObs(y ~ x1)
\end{example}

The variance of the measurement noise $\bm{e}_k$ in (\ref{eq:obs}) is specified per output using \code{\$setVariance}.
\begin{example}
m$setVariance(y ~ s1)
\end{example}
For multiple outputs the covariance between two outputs (e.g \code{y1}, \code{y2}) is specified by
\begin{example}
m$setVariance(y1y2 ~ s12)
\end{example}
\pkg{ctsmr} ensures the variance-covariance matrix is symmetric such that it enough to specify either the lower or upper triangle. If no co-variance between certain output noise are specified they are assumed $0$.

Exogenous inputs must be specified by name by using the \code{\$addInput} method. This tells \pkg{ctsmr} which variables in the specified model are inputs. \pkg{ctsmr} aims at limiting the user as little as possible on how to name states, inputs, outputs and parameters. The name of the states and output are given from the left hand side of the given system and measurement  equations. Knowing the states, inputs and outputs \pkg{ctsmr} assumes that the rest of the variables are parameters.

\subsection{Estimation}
The parameters of the model can either be fixed or estimated. When estimating a parameter an initial value and box constraints must be given. The method \code{\$setParameter()} is used to fix parameters and to set boundaries for the optimization.
\begin{example}
# Fix a = 10
m$setParameter(a = c(init = 10))
# Estimate b: -10 < b < 10
m$setParameter(b = c(init = 0, lower = -10, upper = 10))
\end{example}
When the model is fully specified with parameter values and boundaries it can be fitted to data
\begin{example}
fit <- m$estimate(date = my.data.frame)
\end{example}

\subsection{Implementation}

\pkg{ctsmr} is able to automatically distinguish some linear and non-linear models. This feature is limited due to R not having symbolic algebra capabilities. However, symbolic derivatives can be found using the standard R function \emph{D}. The derivative function is heavily used in \pkg{ctsmr} to separate the drift $\bm{f}\left( \bm{x}_t, \bm{u}_t, t, \bm{\theta} \right) dt$ and diffusion $\bm{\sigma}\left(\bm{u}_t, t, \bm{\theta} \right) d\bm{\omega}_t$ of the SDEs but also when checking for linearity. If the both derivatives of the drift term wrt. the states and inputs respectively does not depend on neither the states nor the inputs then the SDEs are linear. If the same holds for the measurement equations then the entire CTSM is linear. For linear models the coefficients for the states and inputs are extracted using the derivative function. \strong{Note:} This implies that constants in linear models must be parameterized and not directly specified in the model equations. E.g. \code{(a*x1-0.5)*dt} will become \code{a*x1*dt} because the coefficient in front of \code{x1} is found by differentiation. The correct implementation would be \code{(a*x1-b)*dt} and setting \code{b} by call \code{\$setParameter(b = 0.5)}.

\section{Examples}

Three examples are shown here using \pkg{ctsmr}.
\begin{itemize}
   \item A linear model: Continuous time autoregressive (CAR).
   \item A physical non-linear model: Total nitrogen in phytoplankton in Skive fjord.
   \item A medical three compartment model where only one state is observed.
\end{itemize}

\subsection{A linear model: Continuous time autoregressive (CAR)}

This is a simple example of how a continuous time autoregressive (CAR) model is formulated in \pkg{ctsmr}. The Nile dataset from the standard R package \pkg{datasets} contains annual measurements from the river Nile during the period 1871-1970 and is modeled using an Ornstein–Uhlenbeck process for the system equation. The state $x$ is directly observed without noise.
\begin{align*}
   dx &= \theta \left( \beta - x \right) dt + \sigma d\omega \\
    y &= x\ ,
\end{align*}
where $\theta>0$ is the mean reverting rate, $\beta$ the mean value, $\sigma>0$ the volatility.

The CAR model is specified as follows
\begin{example}
library(ctsmr)
# Initialize a CTSM
m.car <- ctsm()
m.car$addSystem(dx ~ theta * (b - x) * dt + exp(sigma)*dw1)
m.car$addObs(y ~ x)
m.car$setVariance(yy ~ exp(S))
\end{example}

The model is complete and printing the model object gives information about the model.
\begin{example}
> m.car
Non-linear state space model with 1 state, 1 output and 0 input

System equations:
  dx1 ~ theta * (b - x1) * dt + exp(sigma) * dw1

measurement equations:
  y ~ x1

No inputs.

Parameters:  x10, b, S, sigma, theta 
\end{example}

This model is identified as a non-linear model and will use the EKF and integrator. Thus the computational time will increase relative to the linear Kalman filter. The SDE can be modified with a dummy input to ensure it will be considered linear.
\begin{example}
# Add a dummy input to the mean b
m.car$addSystem(dx1 ~ theta * (b*dummy - x1) * dt + exp(sigma)*dw1 )
m.car$addInput(dummy)
\end{example}
The model is now considered linear.
\begin{example}
Linear state space model with 1 state, 1 output and 1 input
\end{example}

The initial parameter values and boundaries are set. For the CAR(1) model the initial value of the state \code{x0} and the mean \code{b} are found from data.
\begin{example}
# Set initial values and bounds
m.car$setParameter(x0   = c(init = 1200,0,2000),
               theta = c(init = 1,0,10),
               b     = c(init = 1200, 800, 1500),
               sigma = c(init = 0, -5, 10),
               S     = c(init= -30)
)

# Fit the model
fit <- m$estimate(nile.dat)
\end{example}
The fitted object contains the estimated parameters, uncertainties, data, model and more. Different levels of diagnostic information is available as usual in R, i.e. \code{fit} and \code{summary(fit)}. The \code{summary} function has two optional arguments: \code{correlation} for the correlation structure of the parameter estimates, and \code{extended} which adds information about the derivative of the objective and penalty function for each estimated parameter.

\begin{example}
> summary(fit.lin, extended = TRUE)
Coefficients:
         Estimate  Std. Error     t value    Pr(>|t|)     dF/dPar dPen/dPar
x10    1.1200e+03  1.4388e+02  7.7840e+00  1.3057e-11  3.2326e-07    0.0003
b      9.1342e+02  2.9212e+01  3.1269e+01  0.0000e+00 -2.8857e-06   -0.0053
S     -3.0000e+01          NA          NA          NA          NA        NA
sigma  5.2756e+00  9.6967e-02  5.4406e+01  0.0000e+00  2.4299e-06    0.0002
theta  6.8455e-01  1.6999e-01  4.0270e+00  1.1488e-04 -2.5929e-07    0.0000

Correlation of coefficients:
      x10  b    sigma
b     0.00           
sigma 0.00 0.03      
theta 0.00 0.04 0.69
\end{example}

For comparison a discrete time AR(1) model is fitted to the same dataset.
\begin{example}
> arima(Nile, order = c(1,0,0))

Call:
arima(x = Nile, order = c(1, 0, 0))

Coefficients:
         ar1  intercept
      0.5063   919.5685
s.e.  0.0867    29.1410

sigma^2 estimated as 21125:  log likelihood = -639.95,  aic = 1285.9
\end{example}

The continuous time mean reverting parameter $\theta$ is converted to the corresponding discrete time AR parameter $\exp(-\theta \Delta)=\exp(-0.6845)=0.5043$. A good correspondance between the continuous and discrete time models is observed.

\subsection{A physical non-linear model: Total nitrogen in phytoplankton in Skive fjord}
\begin{figure}
   \centering
   \includegraphics[width=0.9\textwidth]{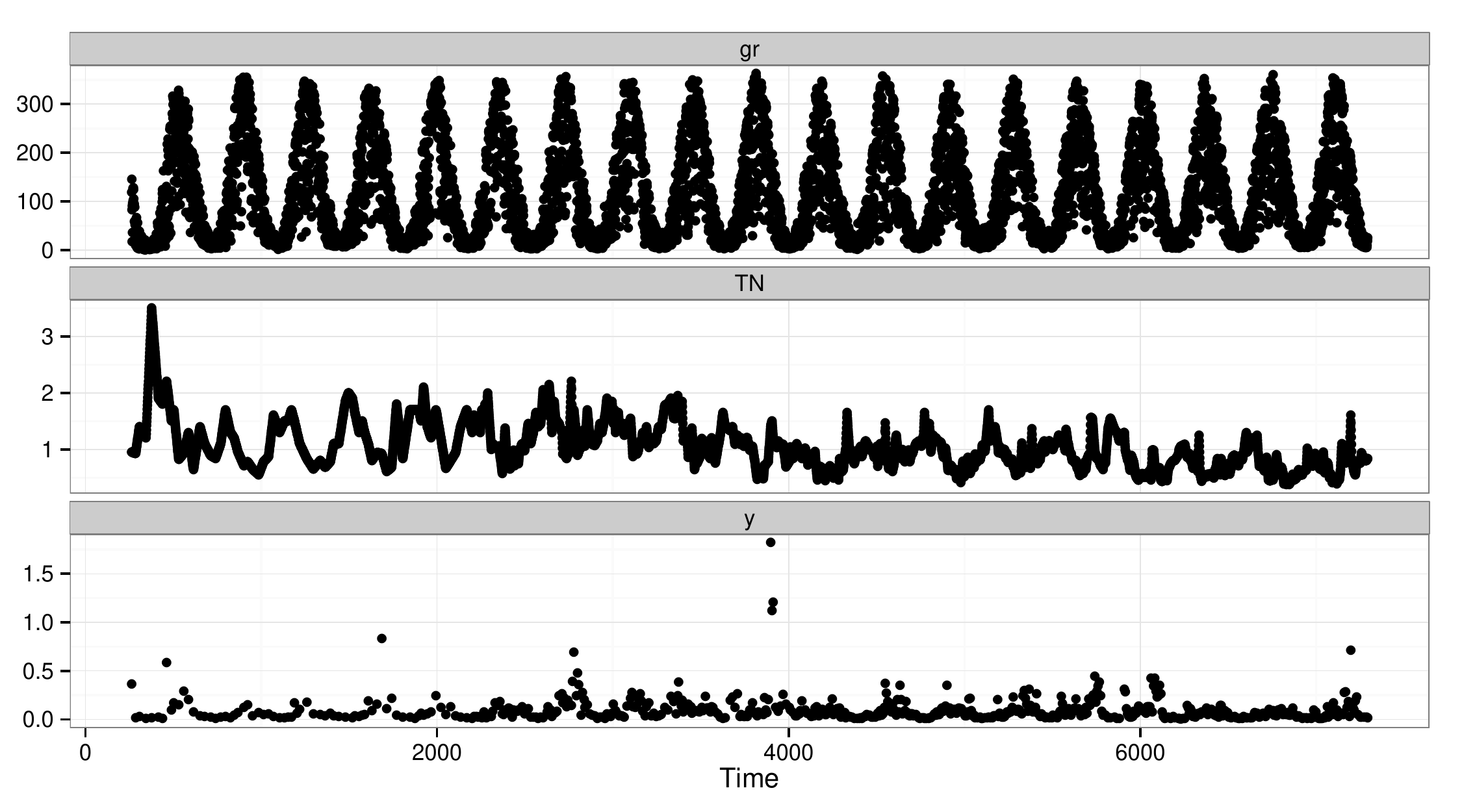}
   \caption{Time series plots of the two inputs and one output: global radiation \texttt{gr}, total nitrogen in the water column \texttt{TN} and total phytoplankton nitrogen \texttt{y}.}\label{fig:skive-data}
\end{figure}

This example is a non-linear ctsmr model to describe the phytoplankton nitrogen dynamics in an estuary located in the northern Denmark \citep{moller_parameter_2011}. The total phytoplankton nitrogen ($X_t$) is described as a function of total nitrogen in the water column ($U_{w,t}$) and incoming global radiation ($U_{gr,t}$) shown in Figure \ref{fig:skive-data}.

The total phytoplankton nitrogen is naturally non-negative and the following linear SDE fulfills that requirement. 
\begin{align}
  dX_{p,t}=(b_0U_{w,t}U_{gr,t}-aX_{p,t})dt+
  \sigma_xX_{p,t}dw_t\label{eq:physde1} ,
\end{align}
Notice that this SDE has a state dependent diffusion term. State dependent diffusion is not well handled by the extended Kalman filter and a Lamperti transformation removing the state dependency is superior. $Z_t$ will be the transformed state by choosing
\begin{align}
   Z_t = \psi(X_t) = \int\frac1{\xi}d\xi\bigg|_{\xi=X_t} = \log \left( X_{p,t} \right)
\end{align}
as the transformation and thus applying Ito's lemma to obtain
\begin{align}
   dZ_t &= \left(\frac{b_0U_{w,t}U_{gr,t}-aX_t}{X_t} -\frac12 \sigma^2 \right) dt + \sigma_x dw_t \nonumber \\
     &= \left(\frac{b_0U_{w,t}U_{gr,t}}{X_t}-a -\frac12 \sigma_x^2 \right) dt + \sigma_x dw_t \nonumber \\
     &= \left( b_0 e^{-Z_t} U_{w,t}U_{gr,t} - a -\frac12 \sigma_x^2 \right) dt + \sigma_x dw_t \label{eq:skive-state-lamperti} \, ,
\end{align}
which is a non-linear SDE describing the same input-output relation with the same parameters, but with state independent diffusion.

The observations are assumed log-normal distributed around the true state. The observation equation
is 
\begin{align}
  \log\left(Y_k\right)=Z_{t_k}+e_{k},\label{eq:modobs}
\end{align} 
where $Y_k$ is the observed nitrogen content in phytoplankton and $e_{k}\sim N(0,\sigma_y^2)$.


The four parameters in (\ref{eq:skive-state-lamperti})-(\ref{eq:modobs}) are all positive and thus are transformed to the entire real axis when implemented in \pkg{ctsmr}.

\begin{example}
> library(ctsmr)
> # New model
> model <- ctsm()
> # The Lamperti transformed system equation
> model$addSystem(dz ~ (exp(lb0-z)*gr*TN - exp(la0) -
>                     0.5*exp(2*lsigma))*dt + exp(lsigma)*dw1)
> # The measurement equation
> model$addObs(ylog~z)
> model$setVariance(ylog~exp(ls0))
> # Define input variables
> model$addInput(gr,TN)

> model$setParameter(z0 = c(init=0,lb=-20,ub=1),
>                    lb0 = c(init=0,lb=-20,ub=1),
>                    la0 = c(init=0,lb=-10,ub=1),
>                    lsigma = c(init=-3,lb=-20,ub=2),
>                    ls0 = c(init=-3,lb=-20,ub=2))
> # Estimate the parameters
> fit <- model$estimate(dat)

> summary(fit)
Coefficients:
         Estimate Std. Error  t value  Pr(>|t|)    
z0      -1.501948   0.436940  -3.4374 0.0006252 ***
la0     -4.058663   0.204437 -19.8529 < 2.2e-16 ***
lb0    -11.011220   0.122263 -90.0617 < 2.2e-16 ***
ls0     -1.647316   0.131516 -12.5256 < 2.2e-16 ***
lsigma  -1.823399   0.086359 -21.1142 < 2.2e-16 ***
---
Signif. codes:  0 ‘***’ 0.001 ‘**’ 0.01 ‘*’ 0.05 ‘.’ 0.1 ‘ ’ 1
\end{example}

\begin{figure}[htbp]
   \centering
   \includegraphics[width = 0.9\textwidth]{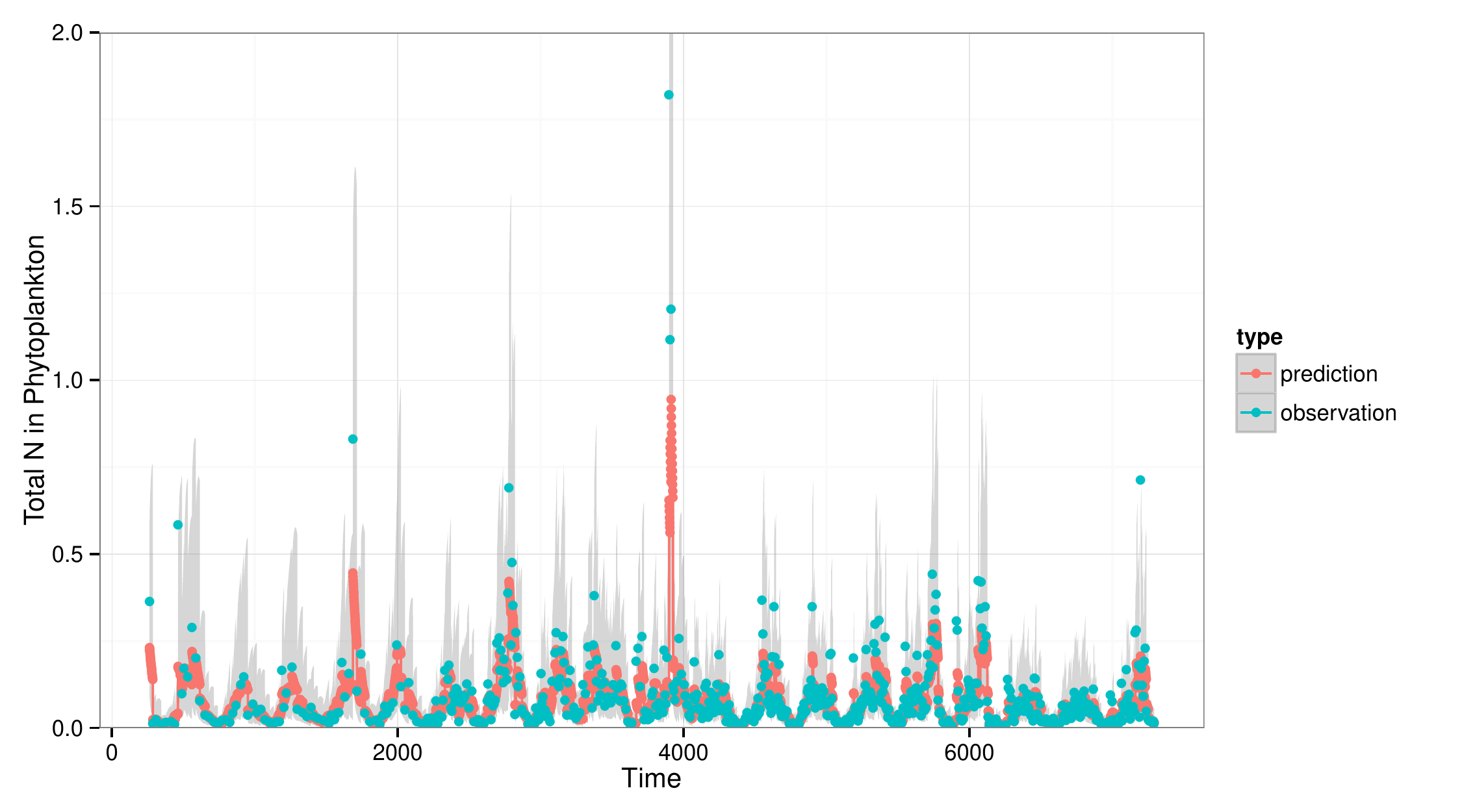}
   \caption{The observed (blue) and predicted (red) total nitrogen in phytoplankton. The red curve is the median in the log-normal distribution.}
   \label{fig:skive_predict}
\end{figure}

The optimization of the log-likelihood converges in 41 iterations and the summary of the estimates shows the estimates, standard errors and test statistics in the usual format in R. The results agree with Table 1 in \citep{moller_parameter_2011}. The one-step predictions are obtained with \code{predict}
\begin{example}
p <- predict(fit)
plot(p, which = "output", engine = "ggplot2", se = TRUE)
\end{example}
The plot of the output and confidence intervals are shown in Figure \ref{fig:skive_predict}.

The model for the total nitrogen in phytoplankton has been further extended, see \citep{kloppenborg:phd:2011}.

\subsection{Three compartment model for insulin}
This example demonstrates a model with more hidden states than observation equations. \citep{linz_papir_2015} simulates a linear 3 compartment model where the three system equations describe the exchange between three compartments each representing a part of the physiological insulin response to consumption of food. Only one of the three states is observed. The response ($y$) is the insulin concentration in the blood of a patient, and the input ($u$) as meals. A diffusion term is added to the first state to account for imperfections in the input $u$. The data is simulated according to the model 
\begin{align}
d\bm{x}_t &= \left(\left[\begin{matrix} u_t\\ 0 \\0
                    \end{matrix}\right]+
   \left[\begin{matrix} -k_a & 0 & 0 \\
                        k_a  & -k_a & 0 \\
                        0    & k_a  & -k_e
                    \end{matrix}\right]\bm{x}_t
    \right)dt+
      \left[\begin{matrix} \sigma_{1} & 0 & 0 \\
                        0  & 0 & 0 \\
                        0    & 0  & 0
                    \end{matrix}\right]d\bm{\omega}_t
   \label{eq:exsde1}\\
   y_k &= \left[\begin{matrix} 0 & 0 & 1\end{matrix}\right]\bm{x}_{t_k} + e_k \, , 
\label{eq:exobs1}
\end{align}
where $\bm{x}\in\mathbb{R}^3$, $\bm{\omega}_t\in\mathbb{R}^3$, $e_k\sim \mathcal N(0,s^2)$, $t_k=\{1,11,21,...\}$ and the specific parameters ($\bm{\theta}$) used for simulation are given in Table~\ref{tab:simex} (first column). All positive parameters have been specified in the log-domain in the \pkg{ctsmr} implementation.

\begin{example}
library(ctsmr)
m3 <- ctsm()
m3$addSystem( dx1 ~ (u-exp(lka)*x1)*dt + exp(lsig1)*dw1 )
m3$addSystem( dx2 ~ (exp(lka)*x1-exp(lka)*x2)*dt )
m3$addSystem( dx3 ~ (exp(lka)*x2-exp(lke)*x3)*dt )
m3$addObs( y ~ x3 )
m3$setVariance( yy ~ exp(lS) )

m3$addInput( u )
# Generate a stochastic realization
sim <- stochastic.simulate(m3, pars = c(
                                 x10   = 40,
                                 x20   = 35,
                                 x30   = 11, 
                                 lka   = log(0.025),
                                 lke   = log(0.08),
                                 lsig1 = log(2),
                                 lS    = log(0.025)), data = u, nsim = 1L)
\end{example}
The \code{stochastic.simulate} generates stochastic realizations of the states and output given a time series of the inputs. The model is prepared for fitting as usual by setting initial values and boundaries.
\begin{example}
m3$setParameter(x10 = c(init=30,0,1000),
                x20 = c(init=30,0,1000),
                x30 = c(init=12,0,100),
                lka  = c(init=-3,-10,3),
                lke  = c(init=-3,-10,3),
                lsig1 = c(init=0,-10,5),
                lS   = c(init=0,-10,5)
                )

# fit model to the simulated data
fit3<-m3$estimate(sim)
\end{example}

Table~\ref{tab:simex} shows the true and estimated parameters side by side. 
\begin{table}
\centering
\caption{\label{tab:simex} True parameter values and estimates for the simulation example, 95\% confidence intervals for the individual parameters are given in parenthesis below the estimates.} 
\begin{tabular}{l|rrrrrrr}
               & $x_{10}$ & $x_{20}$ & $x_{30}$ & $k_a$ & $k_e$ & $\sigma_1$ & $s$ \\
\hline
$\theta$       & 40 & 35 & 11 & 0.025 & 0.08 & 2 & 0.025 \\
$\hat{\theta}$ & 34.492& 36.799 & 10.624 & 0.0249 & 0.0795 & 1.711 & 0.030 \\
$\left[ 2.5\% , 97.5\% \right]$ & \tiny{(17.13,51,85)} & \tiny{(31.62,41,97)} & \tiny{(10.28,10,97)} & \tiny{(0.023,0.027)} & \tiny{(0.073,0.087)} & \tiny{(1.287,2.273)} & \tiny{(0.021,0.044)} \\
\end{tabular}
\end{table}

\subsubsection{Profile likelihood}
The Wald confidence intervals is based on a second order approximation of the likelihood function. 

The profile likelihood of a parameter $\theta_i$ is defined as
\begin{align}
  L_p\left( \theta_i \right) = \max_{\theta \setminus \theta_i} L\left( \theta \right)
\end{align}
which is for a fixed value of $\theta_i$ the likelihood function is maximized for all other parameters. This is easily implemented with \pkg{ctsmr} in a for-loop or using sapply as in the example here.
\begin{example}
which.var <- "lsig1"
var <- seq(Range[1], Range[2], length.out = 20L)

profile_likelihood <- sapply(var, function(x) {
  m4$ParameterValues[which.var, "initial"] <- x
  fit <- m4$estimate(dat)
  fit$loglik
})
\end{example}

\begin{figure}[ht]
  \centering
  \includegraphics[width=\textwidth]{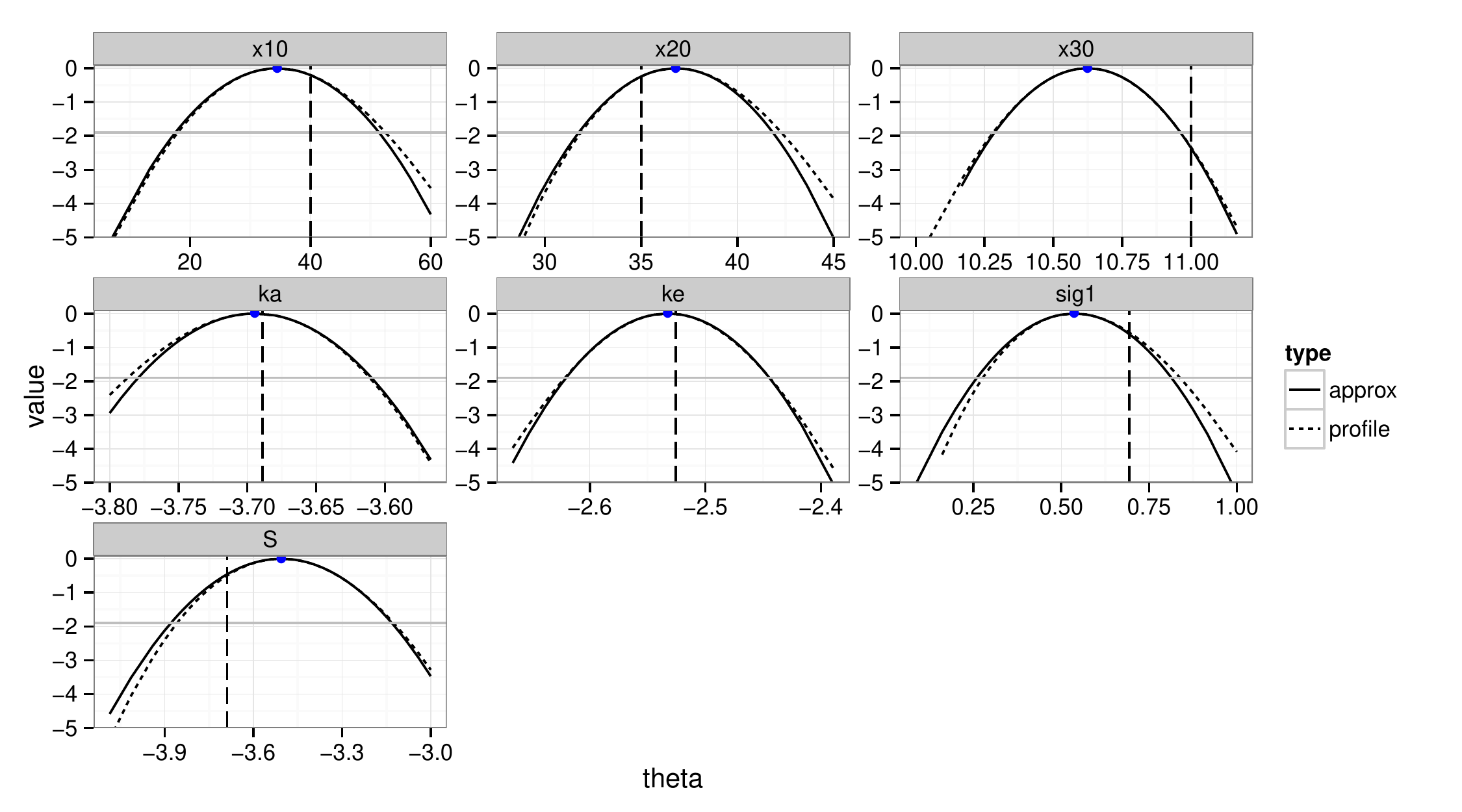}
  \caption{Profile likelihood and Gaussian approximation curves for all parameters. The solid curve is the Wald with the standard deviation from the observed Fisher matrix. The dotted curve is the profile likelihood. The horizontal grey line is the level for the approximate 95\% confidence interval. The blue dot is the estimated parameter value and the vertical dotted line is the true parameter value.}\label{fig:profile_correct}
\end{figure}

The profile likelihood for each parameter in the correct model is shown in Figure~\ref{fig:profile_correct}. The quadratic Wald approximation fits well to the profile likelihood.

If diffusion is added to all three states when estimating the parameters the profile likelihood and Wald approximations deviate significantly for the diffusion and variance parameters.
\begin{figure}[ht]
  \centering
  \includegraphics[width=\textwidth]{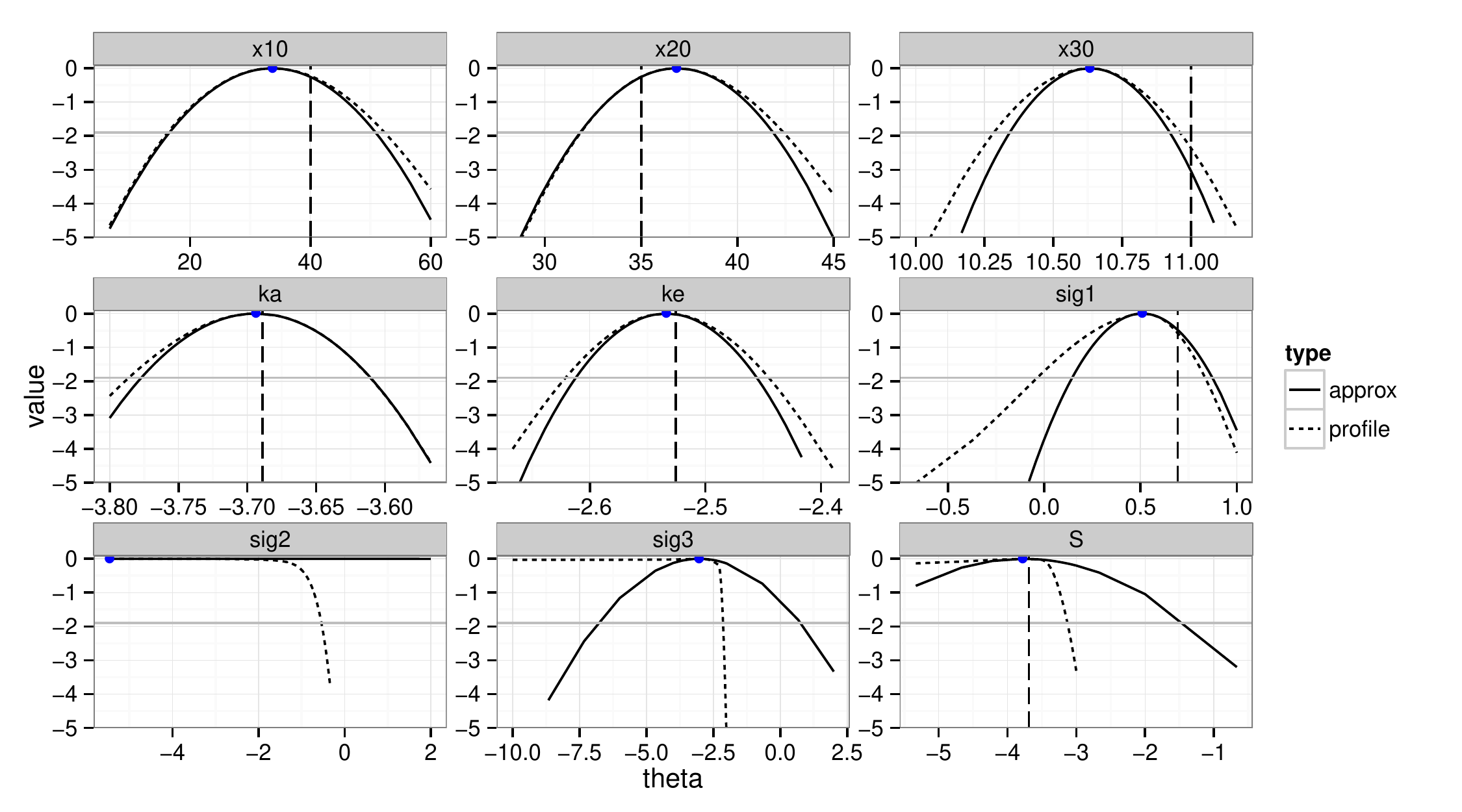}
  \caption{Profile likelihood and Gaussian approximation curves for all parameters. The solid curve is the Wald with the standard deviation from the observed Fisher matrix. The dotted curve is the profile likelihood. The horizontal grey line is the level for the approximate 95\% confidence interval. The blue dot is the estimated parameter value and the vertical dotted line is the true parameter value.}
\end{figure}

\section{Summary}
A general framework for modeling physical dynamical systems using stochastic differential equations has been demonstrated. CTSM-R is an efficient and parallelized implementation in the statistical language R. CTSM-R uses maximum likelihood and thus known techniques for model identification and selection can also be used for this framework.

A detailed user guide as well as additional examples are available from \url{http://ctsm.info}.

\bibliography{juhl-moller-madsen}

\address{Rune Juhl\\
  DTU Compute\\
  Richard Petersens Plads, Building 324, DK-2800 Kgs. Lyngby\\
  Denmark\\}
\email{ruju@dtu.dk}

\address{Jan Kloppenborg Møller\\
  DTU Compute\\
  Richard Petersens Plads, Building 324, DK-2800 Kgs. Lyngby\\
  Denmark\\}
\email{jkmo@dtu.dk}

\address{Henrik Madsen\\
  DTU Compute\\
  Richard Petersens Plads, Building 324, DK-2800 Kgs. Lyngby\\
  Denmark\\}
\email{hmad@dtu.dk}

\end{article}

\end{document}